\documentclass[conference]{IEEEtran}
\IEEEoverridecommandlockouts
\usepackage{cite}
\usepackage{graphicx}
\usepackage{booktabs}
\usepackage{multirow}
\usepackage{hyperref}
\usepackage{amsmath}
\usepackage{array}
\usepackage{caption}
\usepackage{tikz}
\usetikzlibrary{shapes.geometric, positioning, arrows.meta}
\usepackage{tikz}
\usetikzlibrary{shapes.geometric, positioning, arrows.meta}
\usepackage{xcolor}
\usepackage{amssymb}

\begin{document}

\title{Cross-Ecosystem Bug Classification in Quantum Software\thanks{Accepted at \textit{IEEE 9th international conference on computing methodologies and communication. (ICCMC-2026)}, 2026.}}

\author{\IEEEauthorblockN{ Mir Mohammad Yousuf}
\IEEEauthorblockA{\textit{Department of Information Technology} \\
\textit{NIT Srinagar}\\
J\&K, India \\
yousuf\_2022phaite006@nitsri.ac.in}
\and
\IEEEauthorblockN{  Shabir Ahmad Sofi}
\IEEEauthorblockA{\textit{Department of Information Technology} \\
\textit{NIT Srinagar}\\
J\&K, India \\
shabir@nitsri.ac.in}
\and
\IEEEauthorblockN{Bisma Majid}
\IEEEauthorblockA{\textit{Department of Information Technology} \\
\textit{NIT Srinagar}\\
J\&K, India \\
bismabhat\_ite006@nitsri.ac.in}}

\maketitle

\begin{abstract}
Quantum software engineering faces unique challenges due to the interaction of classical and quantum components, which produce complex and often poorly understood bug patterns. Characterizing these bugs is essential for advancing testing, debugging, and quality assurance in quantum ecosystems. This paper presents a comparative study of 12,910 issues from Qiskit and 4,613 issues from 11 additional repositories, including Cirq and PyQuil. Using a rule-based classification framework, we analyze bugs by type, category, severity, quality attributes, and quantum-specific subtypes. Results show that classical bugs consistently dominate ($\sim$67\%) across ecosystems, while quantum-specific bugs account for $\sim$27--30\%. Ecosystem-specific trends emerge: Qiskit repositories exhibit more compatibility-related bugs, whereas other ecosystems show higher syntax and quantum-specific bug rates. Across both ecosystems, gate and circuit issues dominate quantum-specific bugs, though non-Qiskit projects reveal broader diversity, including algorithmic, resource, and hybrid-interface issues. Statistical validation confirms that the framework generalizes at the bug-type level while detecting significant variations at finer levels. Benchmarking against four supervised machine-learning baselines further shows that the rule-based framework consistently outperforms data-driven models, particularly for fine-grained quantum-specific subtypes, while longitudinal analysis (2017-2025) indicates that quantum-specific bugs remain relatively stable over time rather than exhibiting a steady increase. This study provides the first cross-ecosystem comparison of bug distributions in quantum software, demonstrating the utility of an interpretable, automation-ready, rule-based framework for guiding testing, debugging, and quality assurance.
\end{abstract}


\section{Introduction}
Quantum computing has emerged as a disruptive paradigm with the potential to address problems in optimization, chemistry, and cryptography that remain intractable for classical systems~\cite{montanaro2016quantum,preskill2018quantum}. This progress depends on rapidly evolving quantum software ecosystems, which support algorithm design, compilation, simulation, and execution on noisy intermediate-scale quantum (NISQ) hardware.

Developing quantum software poses unique challenges because of its hybrid nature, combining classical control code with quantum-specific operations~\cite{serrano2022quantum}. Consequently, quantum software bugs differ from traditional ones: in addition to functional or compatibility issues, they include quantum-specific errors such as circuit misconfigurations, faulty gate operations, noise, measurement issues, and hybrid quantum-classical integration errors~\cite{zhao2023empirical, zhao2021bugs4q, quetschlich2025experience}. Understanding and categorizing such bugs is essential for advancing software reliability and maintainability.

 Although prior empirical studies have characterized bugs in individual ecosystems~\cite{paltenghi2022bugs, luo2022comprehensive}, three gaps remain. \emph{First}, almost all empirical bug studies in quantum software focus on a single ecosystem (typically Qiskit), leaving open the question of whether observed defect patterns generalize. \emph{Second}, existing classification approaches either use coarse manual taxonomies or apply opaque machine-learning models that require large labeled corpora rarely available in this emerging domain. \emph{Third}, no prior work provides a statistically validated, multi-dimensional comparison of defect distributions across quantum ecosystems together with a benchmark against machine-learning baselines.

 This paper addresses these gaps by (i) validating an interpretable rule-based bug classification framework on a heterogeneous, multi-ecosystem corpus; (ii) quantifying the consistency and divergence of defect patterns across ecosystems through hypothesis-driven statistical tests; (iii) benchmarking the rule-based framework against supervised machine-learning baselines; and (iv) examining longitudinal trends to understand how bug patterns evolve as quantum ecosystems mature.

In earlier work~\cite{yousuf2026bug}, we proposed a rule-based bug classification framework and applied it to 12,910 issues from 36 repositories in the Qiskit ecosystem. That study revealed important insights into bug types, severities, and quality impacts but was restricted to a single ecosystem. This paper extends the framework to 11 additional repositories, analyzing 4,613 issues and comparing them with the original Qiskit dataset.

\subsection{Contributions}
The contributions of this paper are:
\begin{enumerate}
    \item Validation of the rule-based classification framework across multiple quantum software ecosystems, with hypothesis-driven statistical testing using chi-square and Cram\'er's V.
    \item Quantitative benchmarking against four supervised ML baselines, demonstrating the rule-based framework's superiority for fine-grained quantum-specific classification.
    \item Comparative analysis of Qiskit and non-Qiskit repositories, showing both consistent distributions and ecosystem-specific trends.
    \item Fine-grained classification of quantum-specific bugs, revealing dominant types such as gate and circuit errors, as well as algorithmic and resource-level issues.
    \item A longitudinal study of bug evolution from 2017 to 2025, characterizing how defect distributions shift as ecosystems mature.
    \item Practical implications and integration guidelines for embedding the framework in real-world development pipelines (e.g., GitHub Actions, CI workflows).
\end{enumerate}
\section{Background}
Quantum Software Engineering extends classical software engineering principles to quantum computing, where programs manipulate qubits through unitary gates, measurements, and circuit transformations. While conventional bugs (syntax errors, compatibility issues, functional bugs) remain relevant, quantum programming introduces unique categories related to circuit construction, qubit connectivity, algorithm implementation, and hardware constraints.

The current landscape is shaped by NISQ devices~\cite{preskill2018quantum, bharti2022noisy}, which provide tens to hundreds of qubits but suffer from short coherence times, noise, and limited connectivity. Consequently, programs are highly sensitive to low-level hardware characteristics. Bugs may arise from mis-specified gates, invalid topologies, transpilation errors, and resource limitations~\cite{yousuf2026systematic}. These quantum-specific bugs coexist with traditional issues such as poor documentation and integration failures, creating a multi-layered defect space.

\textit{Assumptions and constraints:} Our analysis rests on three assumptions: (i) GitHub issues are a reasonable proxy for the bugs developers encounter; (ii) issue metadata contains sufficient lexical signal for rule-based classification; and (iii) keyword and bigram lexicons curated for Qiskit transfer to other ecosystems sharing the same domain vocabulary. Constraints: we exclude private/industrial repositories and treat each issue as belonging to one dominant class per dimension.

\section{Related Work}
Research in Quantum Software Engineering (QSE) has addressed multiple aspects of quality and correctness. Several works focus on testing and verification, including Floyd--Hoare logic for quantum programs~\cite{ying2012floyd}, model checking for quantum circuits~\cite{ying2021model}, and statistical debugging of quantum programs~\cite{huang2019statistical}. While these methods establish strong theoretical foundations, they do not systematically analyze defect trends in real-world repositories.
 
Empirical studies have only recently gained traction. Researchers~\cite{fingerhuth2018open, li2021understanding} carried out preliminary studies of quantum repositories, identifying documentation and usability challenges. Paltenghi and Pradel~\cite{paltenghi2022bugs} and Luo et al.~\cite{luo2022comprehensive} investigated bug reports in the IBM Qiskit ecosystem, highlighting integration and compatibility as frequent issues. Garcia et al.~\cite{garcia2023quantum} examined testing practices, showing that automated testing remains limited in quantum projects. More recently, Zhao et al.~\cite{zhao2023bugs4q} introduced Bugs4Q, a curated dataset of quantum software bugs.
 
\textit{Bug classification approaches:} Classical bug classification has been extensively studied using rule-based systems, supervised classifiers, and natural language processing techniques~\cite{ahmed2021capbug, bhandari2023buggin}. In quantum software, however, the lack of large labeled corpora makes purely data-driven methods difficult to apply, motivating interpretable rule-based or hybrid approaches.

\textit{Comparison with existing bug-classification work:} Table~\ref{tab:lit_compare} positions our framework against representative classical and quantum bug-classification studies. Compared to classical NLP-based classifiers~\cite{ahmed2021capbug, bhandari2023buggin}, which require large labeled datasets and address only generic dimensions (bug category and priority), our framework operates without supervised training and adds quantum-specific dimensions. Paltenghi and Pradel~\cite{paltenghi2022bugs} is the closest prior cross-platform empirical study and inspects 223 manually labeled bugs across 18 quantum platforms; however, their analysis is fully manual, limited to a single label dimension (recurring bug pattern), and does not provide automated classification or hypothesis-driven statistical comparison. Bugs4Q~\cite{zhao2023bugs4q} curates a reproducible bug benchmark for Qiskit but does not offer a multi-dimensional classifier. To our knowledge, this work is the first to combine automated, interpretable, multi-dimensional classification with statistical cross-ecosystem validation and an ML benchmark.
 
\begin{table}[htbp]
\centering
\caption{Comparison with prior bug-classification work. ``Auto.'' = automated classification, ``Multi-dim.'' = multi-dimensional labels, ``Stat.'' = statistical cross-ecosystem validation.}
\label{tab:lit_compare}
\footnotesize
\begin{tabular}{p{1.3cm}p{0.9cm}p{2.85cm}p{0.45cm}p{0.65cm}p{0.45cm}}
\toprule
\textbf{Ref.} & \textbf{Domain} & \textbf{Method} & \textbf{Auto.} & \textbf{Multi-dim.} & \textbf{Stat.} \\
\midrule
\cite{ahmed2021capbug}      & Classical & ML (NLP)       & \checkmark & Partial & -- \\
\cite{bhandari2023buggin}   & Classical & ML+NLP         & \checkmark & --      & -- \\
\cite{paltenghi2022bugs} & Quantum   & Manual taxonomy (18 platforms) & --        & \checkmark & -- \\
\cite{zhao2023bugs4q}       & Quantum   & Curated benchmark (Qiskit) & --        & --      & -- \\
\cite{yousuf2026bug}        & Quantum   & Rule-based (Qiskit only) & \checkmark & \checkmark & -- \\
\textbf{This work}                  & \textbf{Quantum} & \textbf{Rule-based + ML benchmark (12 ecosystems)} & \textbf{\checkmark} & \textbf{\checkmark} & \textbf{\checkmark} \\
\bottomrule
\end{tabular}
\end{table}
 
In our prior work~\cite{yousuf2026bug}, we proposed a rule-based classification framework and applied it to Qiskit repositories, producing the first structured taxonomy of bug types, severities, and quality attributes in a large quantum ecosystem. The present paper builds on that effort by validating the framework across 11 additional ecosystems (Cirq, PyQuil, Q\#, OpenQL, XACC, Strawberry Fields, Amazon Braket SDK, D-Wave Ocean SDK, Staq, Tequila, Silq), benchmarking against ML baselines, and adding a longitudinal analysis.

\section{Methodology}
\begin{figure}[htbp]
  \centering
  \includegraphics[width=0.65\linewidth]{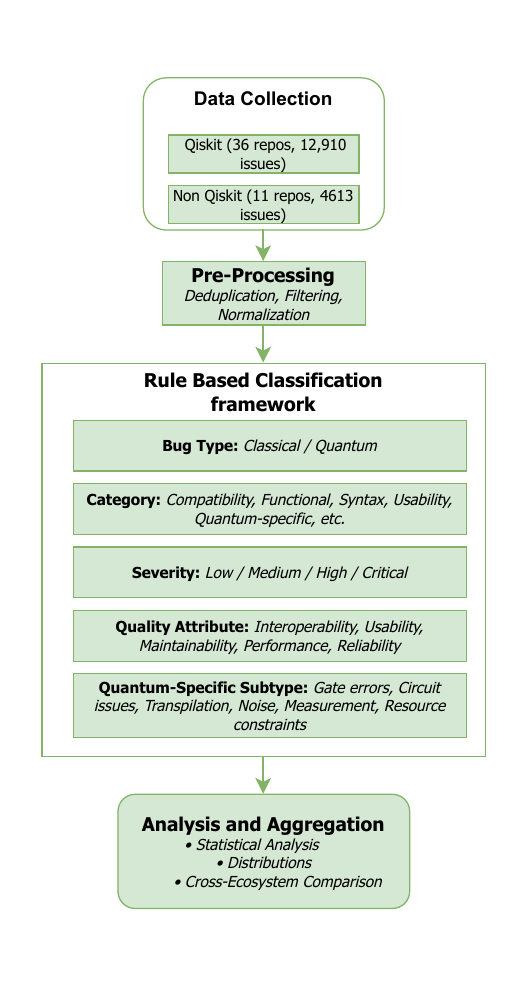}
  \caption{Workflow of the proposed framework:Bug reports from Qiskit (36 repositories, 12,910 issues) and non-Qiskit ecosystems (11 repositories, 4,613 issues) are pre-processed and classified using a rule-based framework across five orthogonal dimensions. Results are aggregated for distributional analysis and cross-ecosystem comparison.}
  \label{fig:methodology}
\end{figure}

To investigate defect characteristics across multiple quantum ecosystems, we adopt a structured rule-based classification framework that analyzes issues along five orthogonal dimensions: bug type, category, severity, quality attribute, and quantum-specific subtype. The framework enables both high-level categorization and fine-grained defect analysis, ensuring comparability between Qiskit and non-Qiskit repositories. As shown in Fig.~\ref{fig:methodology}, the workflow consists of dataset collection, preprocessing, feature extraction, rule-based classification, and analysis.

\subsection{Classification Framework}
The framework, originally established and validated on Qiskit~\cite{yousuf2026bug}, decomposes bugs into five orthogonal dimensions, capturing both general software bugs and those unique to quantum programming:
\begin{itemize}
  \item \textbf{Bug type:} \textit{Classical}, \textit{Quantum}, or \textit{Uncategorized}.
  \item \textbf{Category:} \textit{Compatibility}, \textit{Functional}, \textit{Syntax}, \textit{Usability}, and \textit{Quantum-specific}.
  \item \textbf{Severity:} \textit{Low}, \textit{Medium}, \textit{High}, \textit{Critical} (inferred from labels, keyword weights, and activity-based heuristics).
  \item \textbf{Quality attribute:} ISO/IEC 25010-inspired attributes including \textit{Interoperability}, \textit{Usability}, \textit{Maintainability}, \textit{Performance}, and \textit{Reliability}.
  \item \textbf{Quantum-specific subtype:} \textit{Gate errors}, \textit{Circuit issues}, \textit{Transpilation bugs}, \textit{Noise/measurement bugs}, \textit{Resource constraints}, \textit{Hybrid interface}, \textit{Quantum Error Correction (QEC)}, and \textit{Hardware-specific} errors.
\end{itemize}

\subsection{Preprocessing, Feature Extraction, and Classification}
The pipeline comprises four steps. \emph{Step 1: Preprocessing.} Issue title, description, comments, and labels are merged; markdown, code, and URLs removed; text is lowercased, tokenized, stopwords filtered, and lemmatized. \emph{Step 2: Feature extraction.} Features include (i) weighted keyword lexicons, (ii) weighted bigrams, and (iii) TF--IDF vectors over the corpus and category prototypes, with metadata (labels, comments) aiding severity. \emph{Step 3: Rule application.} Dimensions are computed independently: Bug Type via weighted $\mathrm{argmax}$; Category via labels and keywords; Severity via labels, keywords, and comment thresholds (5, 10), with security terms forcing \emph{Critical}; Quality Attribute via keyword and TF--IDF similarity (priority-based tie-breaking); Quantum subtype via combined keyword–TF--IDF scoring ($\lambda=2.0$). \emph{Step 4: Output.} Each issue is assigned a five-dimensional label.

\subsection{Datasets}
Two datasets are used:
\begin{itemize}
  \item \textbf{Qiskit dataset:} 12,910 issues from 36 repositories in the Qiskit ecosystem~\cite{qiskit2021}, as reported in our earlier work~\cite{yousuf2026bug}.
  \item \textbf{Non-Qiskit dataset:} 4,613 issues collected from 11 repositories: Cirq~\cite{cirq}, PyQuil~\cite{pyquil}, Q\#~\cite{qsharp}, OpenQL~\cite{openql}, XACC~\cite{xacc}, Strawberry Fields, Amazon Braket SDK, D-Wave Ocean SDK~\cite{dwave_ocean}, Staq~\cite{staq}, Tequila~\cite{tequila}, and Silq~\cite{silq}. The repository-wise distribution is given in Table~\ref{tab:repos}.
\end{itemize}

All issues underwent identical preprocessing: deduplication, removal of non-bug entries (feature requests, discussions), and text/metadata normalization, ensuring methodological consistency between ecosystems. Following classification, labeled issues are aggregated to analyze distributions of bug types, categories, severities, and quality attributes. Quantum-specific subtypes are examined separately. Non-Qiskit results are then compared against the earlier Qiskit dataset, enabling cross-ecosystem study of commonalities and divergences.

\begin{table}[htbp]
\centering
\caption{Repositories and issue counts (non-Qiskit dataset)}
\label{tab:repos}
\begin{tabular}{lc}
\hline
\textbf{Repository} & \textbf{Number of Issues} \\
\hline
Cirq                        & 2316 \\
PyQuil                      & 784  \\
Q\#                         & 558  \\
OpenQL                      & 276  \\
XACC                        & 241  \\
Strawberry Fields           & 139  \\
Amazon Braket SDK (Python)  & 118  \\
D-Wave Ocean SDK            & 73   \\
Staq                        & 47   \\
Tequila                     & 32   \\
Silq                        & 29   \\
\hline
\textbf{Total}              & \textbf{4613} \\
\hline
\end{tabular}
\end{table}

\subsection{Statistical Analysis}
To validate the generalizability of the framework, we conduct hypothesis-driven statistical tests across all five dimensions. For each dimension, the null hypothesis ($H_{0}$) is that the bug distribution is independent of the ecosystem (Qiskit vs.\ non-Qiskit), with the alternative ($H_{1}$) being that distributions differ. Chi-square tests of independence are applied to contingency tables. Statistical significance is assessed at $\alpha=0.05$, and Cram\'er's V quantifies effect size, interpreted as negligible ($V<0.1$), small ($0.1 \leq V < 0.3$), medium ($0.3 \leq V < 0.5$), or large ($V \geq 0.5$).

\subsection{Comparison with Machine-Learning Baselines}
To complement statistical validation, we benchmark the rule-based framework against four supervised classifiers: Logistic Regression, Decision Tree, Random Forest, and Gradient Boosting. All baselines are trained on TF--IDF features extracted from the same preprocessed text, using a stratified 80/20 train--test split on the manually annotated subset described in our earlier work~\cite{yousuf2026bug}. Hyperparameters are tuned via 5-fold cross-validation. Macro-averaged F1 is the primary metric, ensuring fair treatment of minority classes such as \textit{Quantum Noise Issues} or \textit{QEC Bugs}.

\section{Results and Findings}

\subsection{Bug Type Distribution}
Across the 11 non-Qiskit repositories, 65\% of bugs were classical and 31\% were quantum-specific, with 5\% uncategorized. These proportions are nearly identical to the Qiskit dataset (67\% classical, 27\% quantum). This consistency indicates that while classical issues dominate, quantum-specific bugs remain a significant portion across ecosystems.  

\begin{table}[htbp]
\centering
\caption{Bug Type Distribution}
\begin{tabular}{lcc}
\hline
\textbf{Bug Type} & \textbf{Qiskit} & \textbf{Other Repos} \\
\hline
Classical Bugs   & 67\% & 65\% \\
Quantum Bugs     & 27\% & 31\% \\
Uncategorized    & 6\%  & 5\%  \\
\hline
\end{tabular}
\end{table}

\subsubsection{Temporal Evolution of Bugs}
\begin{figure}[t]
    \centering
    \includegraphics[width=0.85\linewidth]{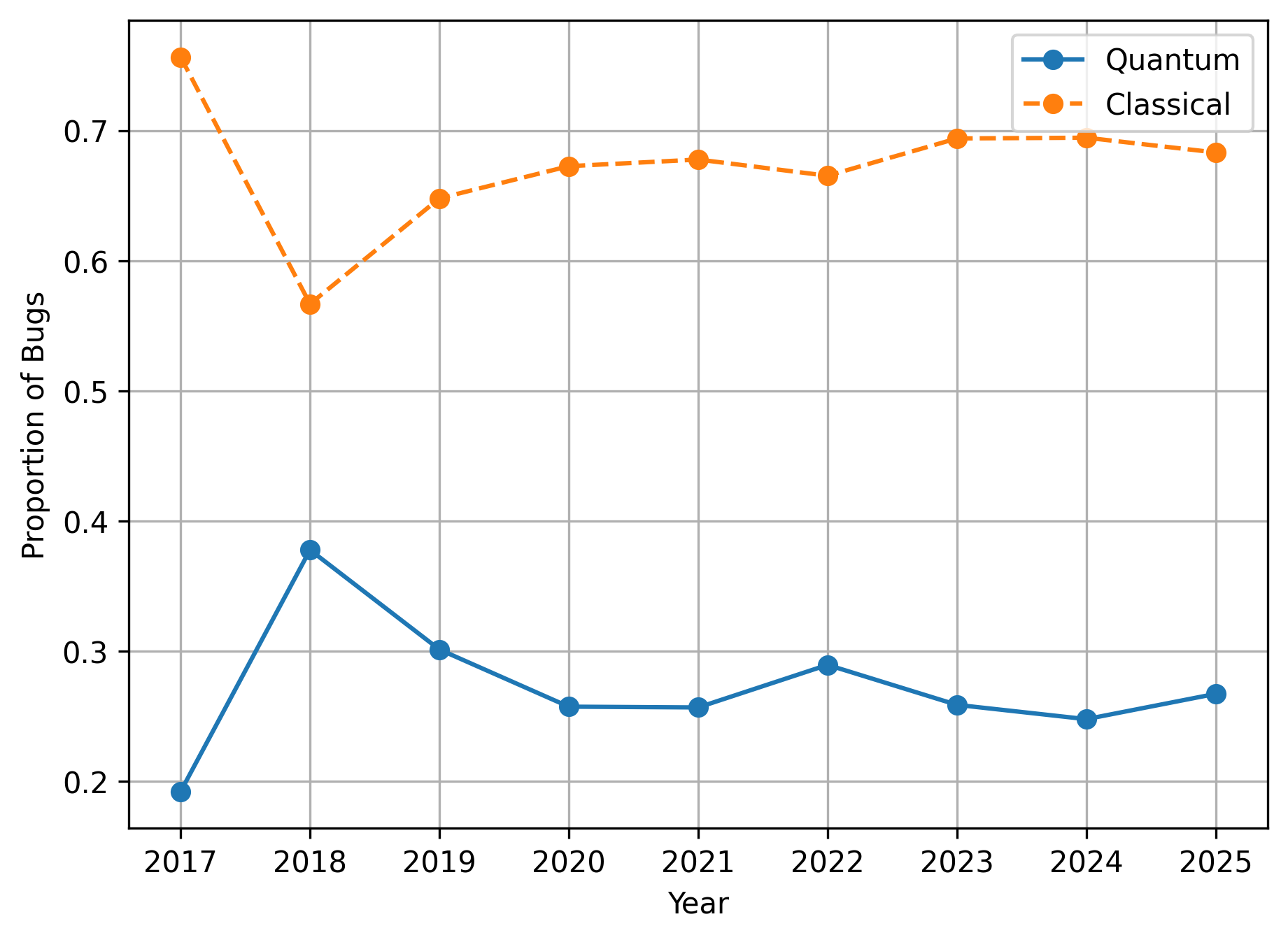}
    \caption{Temporal evolution of quantum and classical bugs in Qiskit (2017--2025). Classical bugs dominate throughout the period, while quantum bugs exhibit higher variability with an early peak followed by stabilization.}
    \label{fig:qiskit_temporal}
\end{figure}

\begin{figure}[t]
    \centering
    \includegraphics[width=0.85\linewidth]{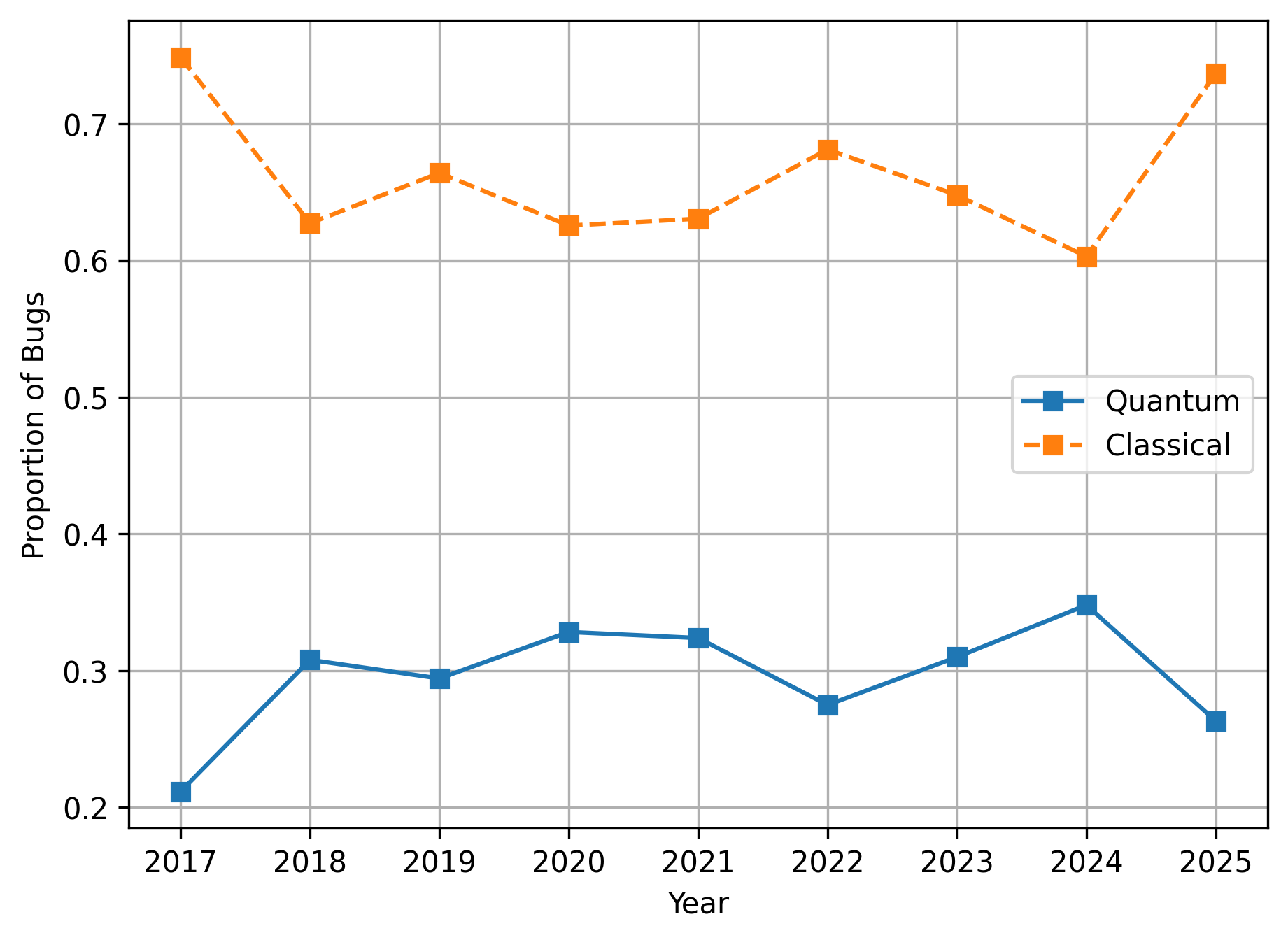}
    \caption{Temporal evolution of quantum and classical bugs in non-Qiskit repositories. The proportions remain relatively stable over time, with classical bugs consistently dominating and quantum bugs showing moderate fluctuations.}
    \label{fig:others_temporal}
\end{figure}

We examine how bug distributions evolve over time. Fig.~\ref{fig:qiskit_temporal} and Fig.~\ref{fig:others_temporal} plot the yearly share of quantum and classical bugs between 2017 and 2025 for Qiskit and other repositories, respectively. Across both datasets, classical bugs consistently dominate, accounting for approximately 60--75\% of reported issues, while quantum bugs remain within 25--40\%.

Rather than exhibiting a monotonic increase, the proportion of quantum bugs fluctuates over time. In Qiskit, quantum bugs peak around 2018 ($\sim$38\%) and subsequently stabilize near 25--30\%. Other repositories show more gradual and stable behavior, with quantum bugs remaining relatively consistent across years.

These trends suggest that quantum software ecosystems have not yet transitioned to predominantly domain-specific bugs. Instead, classical software engineering challenges continue to play a central role, while quantum-specific bugs persist as a consistently secondary category with moderate fluctuations over time.

\subsection{Bug Categories}
The most common categories in non-Qiskit repositories were compatibility (17\%), quantum-specific issues (16\%), functional (15\%), and syntax errors (15\%). In Qiskit repositories, compatibility (25\%) and functional (19\%) dominated, while quantum-specific (12\%) and syntax (10\%) were proportionally lower.  

\begin{figure}
    \centering
    \includegraphics[width=0.99\linewidth]{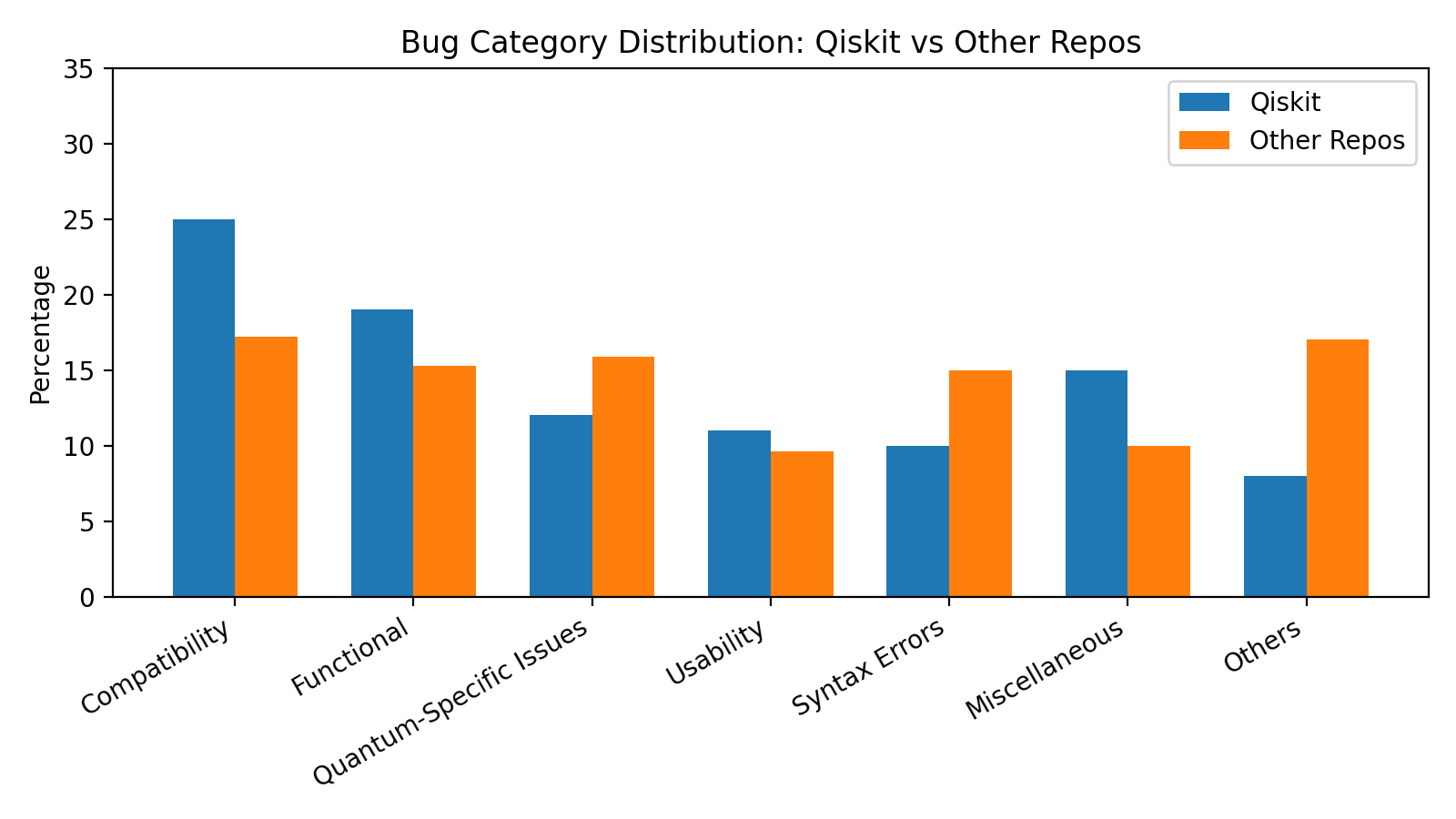}
    \caption{Comparison of bug categories between Qiskit repositories (36 repos, 12,910 issues) and non-Qiskit repositories (11 repos, 4,613 issues). Qiskit projects show a higher share of compatibility bugs, while non-Qiskit projects exhibit more syntax and quantum-specific issues.}
    \label{fig:placeholder}
\end{figure}

\subsection{Severity Distribution}
Both datasets show bugs are dominated by low-severity issues (69\% in Qiskit vs. 72\% in other repositories). However, critical bugs are slightly higher in non-Qiskit projects (24\% vs. 21\%). Medium severity is less frequent outside Qiskit (4\% vs. 7.5\%), and high severity is rare in both.  

\begin{table}[htbp]
\centering
\caption{Severity Distribution}
\begin{tabular}{lcc}
\hline
\textbf{Severity} & \textbf{Qiskit (\%)} & \textbf{Other Repos (\%)} \\
\hline
Low      & 69   & 72 \\
Critical & 21   & 24 \\
Medium   & 7.5  & 4  \\
High     & 2    & $\sim$0 \\
\hline
\end{tabular}
\end{table}

\subsection{Quality Attributes Impacted}
Bugs most frequently impacted interoperability (27\%) and usability (20\%) in other repositories, consistent with Qiskit (24\% and 20\%, respectively). However, Qiskit projects show a stronger focus on maintainability (15\% vs. 9\%), while other repositories show higher reliability and miscellaneous quality concerns.  

\begin{figure}
    \centering
    \includegraphics[width=0.99\linewidth]{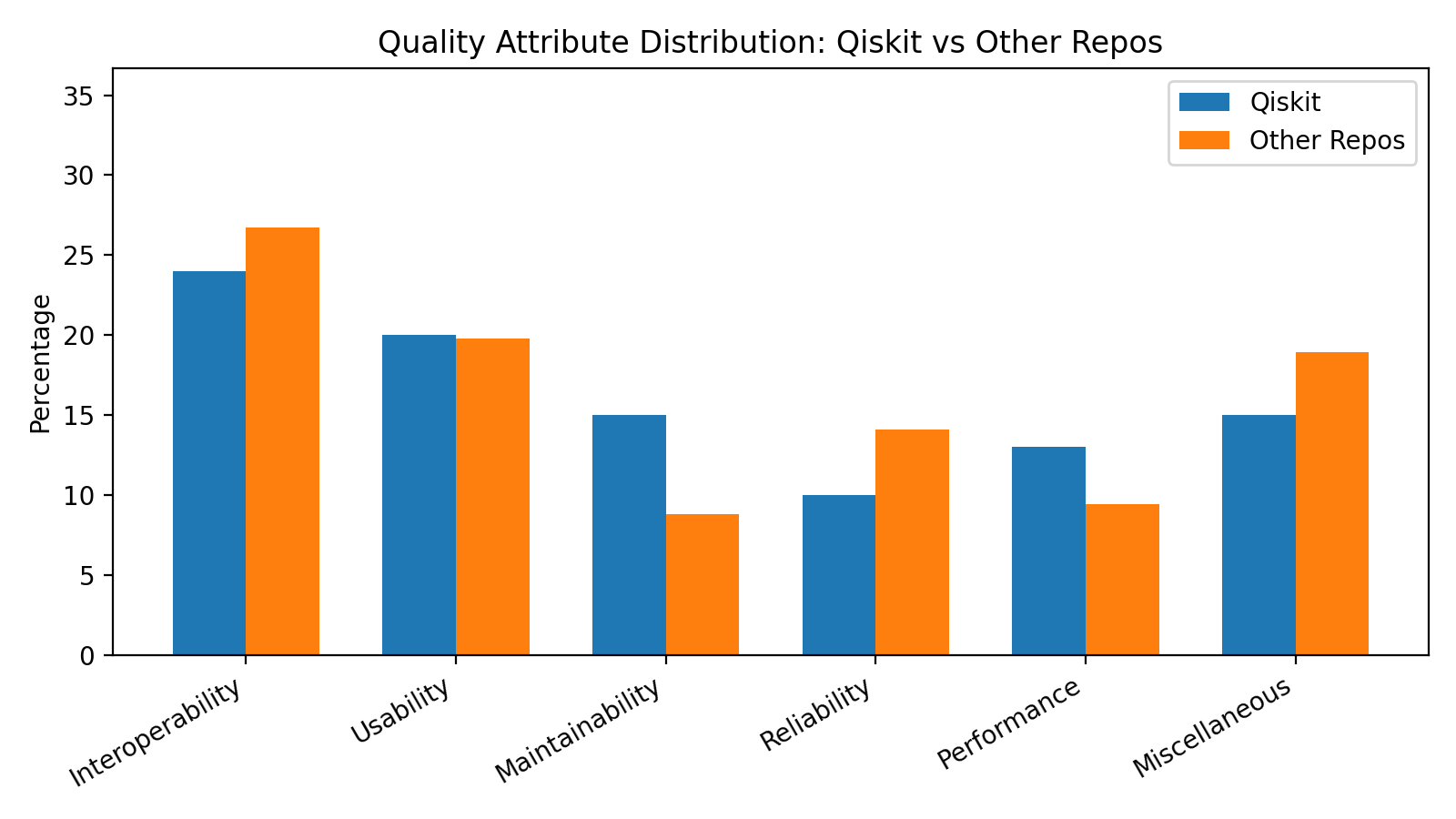}
    \caption{Comparison of quality attributes impacted by bugs in Qiskit and non-Qiskit repositories. Interoperability and usability dominate in both ecosystems, but Qiskit shows more maintainability concerns, while non-Qiskit projects emphasize reliability and miscellaneous issues.}
    \label{fig:placeholder}
\end{figure}

\subsection{Quantum-Specific Bug Types}
A detailed breakdown of quantum-specific issues reveals that circuit and gate errors dominate across both ecosystems.  
\begin{itemize}
    \item Qiskit: Circuit (32\%), Gate (27\%), Transpilation (12\%), Noise (7\%), Measurement (5\%).
    \item Other Repos: Gate (31\%), Circuit (30\%), Transpilation (6\%), Algorithm (6\%), Resource constraints (5\%), plus noise/measurement/decoherence (12\%).
\end{itemize}
While both ecosystems exhibit strong prevalence of circuit- and gate-level bugs ($\sim$60\%), non-Qiskit repositories highlight additional algorithmic and resource-level challenges not as prominent in Qiskit. 

\begin{figure}
    \centering
    \includegraphics[width=0.99\linewidth]{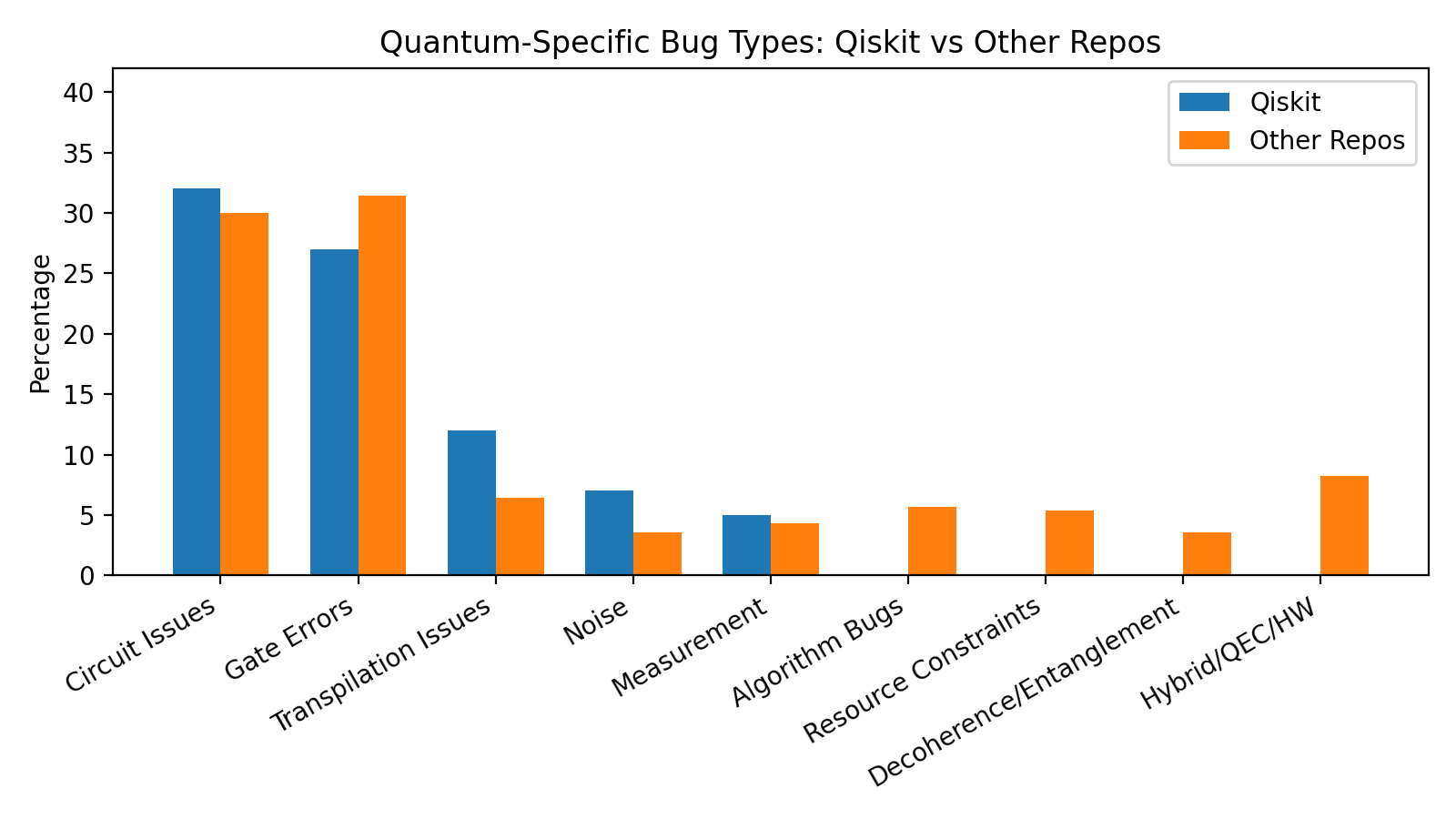}
    \caption{Comparison of fine-grained quantum-specific bug types across ecosystems. Circuit and gate-level bugs dominate both Qiskit and non-Qiskit repositories, but non-Qiskit projects also highlight algorithmic and resource constraint bugs that are less prominent in Qiskit.}
    \label{fig:placeholder}
\end{figure}

\subsection{Comparison with Machine-Learning Baselines}
Table~\ref{tab:ml_baselines} reports macro-F1 of the four supervised baselines compared with the rule-based framework on the manually annotated subset~\cite{yousuf2026bug}. The rule-based framework consistently outperforms all baselines across every dimension. The largest gains appear for fine-grained, low-frequency tasks: Bug Category (0.69 vs.\ 0.26 best baseline) and Quantum-Specific Subtype (0.77 vs.\ 0.15). These results confirm that, in the data-scarce, semantically specialized quantum-software domain, knowledge-driven heuristics outperform shallow data-driven classifiers while remaining interpretable.

\begin{table}[htbp]
\centering
\caption{Macro-F1 of ML baselines vs.\ rule-based framework (Type=Bug Type, Sev.=Severity, Cat.=Category, Qual.=Quality, Q-Sub.=Quantum-specific Subtype.}
\label{tab:ml_baselines}
\footnotesize
\begin{tabular}{lccccc}
\hline
\textbf{Model} & \textbf{Type} & \textbf{Sev.} & \textbf{Cat.} & \textbf{Qual.} & \textbf{Q-Sub.} \\
\hline
Logistic Regression  & 0.45 & 0.43 & 0.25 & 0.39 & 0.11 \\
Decision Tree        & 0.44 & 0.50 & 0.17 & 0.40 & 0.24 \\
Random Forest        & 0.45 & 0.34 & 0.18 & 0.38 & 0.11 \\
Gradient Boosting    & 0.48 & 0.53 & 0.26 & 0.50 & 0.15 \\
\textbf{Rule-based}  & \textbf{0.75} & \textbf{0.68} & \textbf{0.69} & \textbf{0.76} & \textbf{0.77} \\
\hline
\end{tabular}
\end{table}

\subsection{Statistical Validation}

Table~\ref{tab:stat_validation} reports the results of chi-square tests across all bug dimensions. 

Bug type distributions show no significant difference ($\chi^{2}=4.75$, $p=0.093$, $V=0.019$). We \textit{fail to reject $H_{0}$}, confirming that the framework generalizes at the highest level: both Qiskit and non-Qiskit ecosystems contain $\sim$67\% classical and $\sim$27--30\% quantum bugs.

Bug categories show significant differences ($\chi^{2}=277.84$, $p<0.0001$, $V=0.126$). We \textit{reject $H_{0}$} with a small effect size. Qiskit repositories report proportionally more compatibility bugs, while non-Qiskit repositories show higher syntax and quantum-specific bug rates.

Severity distributions differ significantly as well ($\chi^{2}=151.95$, $p<0.0001$, $V=0.093$). We \textit{reject $H_{0}$}, though the effect size is small. Differences likely stem from heterogeneous labeling and reporting practices rather than intrinsic bug characteristics.

Quality attributes also show small but significant differences ($\chi^{2}=214.39$, $p<0.0001$, $V=0.111$). We \textit{reject $H_{0}$}, with interoperability and usability dominating both ecosystems but in slightly different proportions.

Quantum-specific subtypes exhibit the strongest divergence ($\chi^{2}=94.33$, $p<0.0001$, $V=0.227$). We \textit{reject $H_{0}$} with a moderate effect size. Qiskit bugs are dominated by gate and circuit issues, whereas non-Qiskit repositories display a broader distribution that includes algorithmic, decoherence, resource-related, and hybrid-interface bugs.

\begin{table}[htbp]
\centering
\caption{Statistical validation of Qiskit vs.\ non-Qiskit bug distributions (rule-based classification)}
\label{tab:stat_validation}
\begin{tabular}{lccc}
\hline
\textbf{Dimension} & \textbf{$\chi^{2}$} & \textbf{$p$-value} & \textbf{Cram\'er’s V} \\
\hline
Bug Type                 & 4.75    & 0.093    & 0.019 (ns/negligible) \\
Bug Category             & 277.84  & $<$0.0001 & 0.126  \\
Severity                 & 151.95  & $<$0.0001 & 0.093 \\
Quality Attribute        & 214.39  & $<$0.0001 & 0.111  \\
Quantum-Specific Subtype & 94.33   & $<$0.0001 & 0.227 \\
\hline
\end{tabular}
\end{table}

\subsection{Key Findings}

The validation results highlight both framework generalization and ecosystem-specific differences.

\textbf{Generalization:}
\begin{itemize}
    \item The framework generalizes strongly at the bug-type level: no significant difference was found between Qiskit and non-Qiskit repositories. 
    \item Classical bugs consistently dominate ($\sim$67\%) while quantum-specific bugs account for $\sim$27--30\% across ecosystems. 
    \item Interoperability and usability emerge as the most frequent quality attributes in both ecosystems, indicating shared quality concerns.
\end{itemize}

\textbf{Ecosystem-Specific Differences:}
\begin{itemize}
    \item At the bug-category level, Qiskit repositories show more compatibility bugs, while non-Qiskit repositories show higher syntax and quantum-specific bug rates. 
    \item Severity distributions differ significantly but with small effect sizes, reflecting heterogeneous labeling conventions across ecosystems. 
    \item Quantum-specific subtypes diverge most strongly: Qiskit bugs are dominated by gate and circuit issues, whereas non-Qiskit repositories show broader diversity, including algorithmic, decoherence, resource-related, and hybrid-interface bugs.
\end{itemize}

Thus, we \textit{fail to reject $H_{0}$ at the bug-type level}, confirming framework generalization, while \textit{rejecting $H_{0}$ for all finer dimensions}, demonstrating that the framework is robust yet sensitive enough to reveal ecosystem-specific bug patterns. This balance of stability and sensitivity makes the rule-based framework both interpretable and actionable for guiding testing, debugging, and quality assurance in quantum software engineering.

\section{Discussion}
The comparative analysis shows that the framework achieves both stability and sensitivity. Bug-type distributions are statistically consistent across ecosystems (fail to reject $H_{0}$), confirming robust generalization. At finer levels, $H_{0}$ is rejected for categories, severity, quality attributes, and quantum-specific subtypes, revealing ecosystem-specific patterns. Qiskit reports more compatibility bugs, reflecting integration complexity and API evolution; non-Qiskit shows higher syntax and quantum-specific bug rates linked to diverse compilers and experimental language features. The most pronounced divergence is in quantum-specific subtypes (moderate effect size), suggesting ecosystem-tailored strategies: Qiskit projects benefit from improved integration testing and API stability, while other platforms require stronger syntax validation, compiler robustness, and resource-aware tooling.

Temporal analysis shows that classical bugs dominate across both ecosystems, with no clear rise in quantum-specific bugs. Quantum bugs remain stable over time, indicating that development is still largely shaped by traditional software engineering challenges rather than a shift toward domain-specific issues.

The framework integrates into workflows via: (i) a \emph{GitHub Action} that labels new issues by type, severity, and quality attribute; (ii) a \emph{CI/CD plug-in} (e.g., GitLab CI, Jenkins) that generates periodic issue analytics dashboards; and (iii) a \emph{pre-commit linter} that flags commits matching high-severity quantum-specific patterns.

As shown in Table~\ref{tab:ml_baselines}, performance gains stem from: (i) sparse labels for fine-grained classes (e.g., QEC bugs), (ii) semantic overlap between similar categories, and (iii) limitations of TF--IDF models relying on surface features. The rule-based approach encodes expert knowledge, enabling better generalization without large labeled datasets, especially in emerging ecosystems.

\section{Threats to Validity}
\textbf{Internal:} the rule-based framework may misclassify ambiguous reports; we mitigate this by applying consistent rules across ecosystems and validating outputs against a manually annotated benchmark~\cite{yousuf2026bug}. \textbf{Construct:} severity labels are inconsistently applied, and quality-attribute mappings rely on prototype documents that may introduce bias; conservative fallback to \emph{Miscellaneous} reduces forced misclassifications. \textbf{External:} while our dataset covers 36 Qiskit and 11 additional repositories spanning major SDKs, results may not fully generalize to industrial or proprietary projects. \textbf{Statistical:} chi-square tests can be sensitive to small category counts; we focus on aggregate distributions and report Cram\'er's V to interpret effect sizes meaningfully.

\section{Conclusion and Future Work}This paper validated and extended a rule-based bug classification framework, originally developed for Qiskit, by applying it to 11 additional quantum software repositories spanning ecosystems such as Cirq, PyQuil, Q\#, OpenQL, and Braket. An analysis of 4,613 non-Qiskit issues alongside 12,910 Qiskit issues confirmed that the framework generalizes strongly at the bug-type level ($p = 0.093$, fail to reject $H_{0}$), while statistical validation revealed significant ecosystem-specific differences in bug categories, severity, quality attributes, and quantum-specific subtypes. 
Benchmarking against four machine-learning baselines (Logistic Regression, Decision Tree, Random Forest, Gradient Boosting) showed that the rule-based framework consistently outperforms them, by up to 0.62 macro-F1 on quantum-specific subtypes, while remaining interpretable.
A longitudinal analysis (2017--2025) further indicated that classical bugs dominate across ecosystems, with quantum bugs remaining relatively stable over time, highlighting the continued prominence of traditional software engineering challenges alongside domain-specific issues.
Building on these findings, future work will pursue hybrid pipelines that combine rule-based heuristics with large language models to improve recall on rare categories, develop standardized severity metrics across quantum projects to enable fair longitudinal comparison, extend the framework to industrial and closed-source ecosystems for stronger external validity, and integrate it more tightly with developer tooling such as IDE plug-ins, GitHub Actions, and CI dashboards so it can directly support real-time issue triage and quality assurance in production quantum software development.

\bibliographystyle{IEEEtran}
\bibliography{references}

\end{document}